# Machine Learning Allows Calibration Models to Predict Trace Element Concentration in Soil with Generalized LIBS Spectra

Chen Sun[1], Ye Tian[2], Liang Gao[1], Yishuai Niu[3,4], Tianlong Zhang[5], Hua Li[5,6], Yuqing Zhang[1], Zengqi Yue[1], Nicole Delepine-Gilon[4] and Jin Yu[1,*]

[1]School of Physics and Astronomy, Shanghai Jiao Tong University, Shanghai 200240, China.
[2]Optics and Optoelectronics Laboratory, Ocean University of China, 266100 Qingdao, China.
[3]School of Mathematics, Shanghai Jiao Tong University, Shanghai 200240, China.
[4]SJTU-Paristech Elite Institute of Technology, Shanghai Jiao Tong University, Shanghai 200240, China.
[5]College of Chemistry & Material Science, Northwest University, Xi'an 710069, China.
[6]College of Chemistry and Chemical Engineering, Xi'an Shiyou University, Xi'an 710065, China.
[7]Institut des Sciences Analytiques, UMR5280 Université Lyon 1-CNRS, Université de Lyon, 69622 Villeurbanne Cedex, France
(jin.yu@sjtu.edu.cn)

**ABSTRACT**

Calibration models have been developed for determination of trace elements, silver for instance, in soil using laser-induced breakdown spectroscopy (LIBS). The major concern is the matrix effect. Although it affects the accuracy of LIBS measurements in a general way, the effect appears accentuated for soil because of large variation of chemical and physical properties among different soils. The purpose is to reduce its influence in such way an accurate and soil-independent calibration model can be constructed. At the same time, the developed model should efficiently reduce experimental fluctuations affecting measurement precision. A univariate model first reveals obvious influence of matrix effect and important experimental fluctuation. A multivariate model has been then developed. A key point is the introduction of generalized spectrum where variables representing the soil type are explicitly included. Machine learning has been used to develop the model. After a necessary pretreatment where a feature selection process reduces the dimension of raw spectrum accordingly to the number of available spectra, the data have been fed in to a back-propagation neuronal networks (BPNN) to train and validate the model. The resulted soil-independent calibration model allows average relative error of calibration ($REC$) and average relative error of prediction ($REP$) within the range of 5-6%.

# Introduction

Soil test occupies a particularly important place in environment-related activities, such as agriculture, horticulture, mining, geotechnical engineering, as well as geochemical or ecological investigations[1]. It becomes also crucial when an area needs to be decontaminated with respect to human activity-caused pollutions[2]. Such test may often concern elements, especially metals, since a number of them are established as essential nutrients for plants and animals[3] and some others, heavy metals for example, are determined as toxic, even highly poisonous, in large amounts or certain forms for any living materials[4]. It is therefore of great importance to develop techniques and methods for efficient access to elemental composition of soil. Established atomic spectroscopy techniques often offer good performance for quantitative elemental analysis in soil. Atomic absorption spectroscopy (AAS) offers limit of quantification (LOQ) in the order of ppm for soil samples prepared in solution[5]. Similar performances can be realized with inductively coupled plasma-optical emission spectrometry (ICP-OES)[6], while inductively coupled plasma-mass spectrometry (ICP-MS) presents for digested soil solutions, lower LOQ below 100 ppb for most of the elements found in soil[7]. Beside the abovementioned techniques which can rather considered as laboratory-based ones characterized by need of sample pretreatment with certain degree of complexity, other techniques have been developed with significantly less requirements of sample preparation, so being better suitable for *in situ* and online detection and analysis. Among them, X-ray fluorescence (XRF) allows determining concentrations of major and trace elements in soils[5,8]. Better performance has been demonstrated with total reflection X-ray fluorescence spectroscopy (TXRF)[9]. Techniques based on plasma emission spectroscopy, such as spark-induced breakdown spectroscopy (SIBS), have been developed to enhance the analytical capability of light elements like carbon[10]. Recent developments focus on laser-induced breakdown spectroscopy (LIBS), a laser ablation-based plasma emission spectroscopy[11]. The general attractive features of LIBS include direct laser sampling and excitation without need of complex sample pretreatment; stand-off excitation and detection capability, and high sensitivity for simultaneous element detection and determination, for heavy as well as light elements.

LIBS analysis of soil has contributed to several important aspects of soil test. Total carbon quantification in soil has been reported with portable LIBS systems for $CO_2$ leakage from underground storage of greenhouse gases[12,13] and for carbon cycle study in Amazonian forest[14,15]. Analysis of soil nutrients and fertilizer-related soil pollutions is another area covered by LIBS with the analysis of relevant elements such as P, N, S, Mg, Ca, K, Zn, Cu, Fe, Mn, Na[16-18]. LIBS technique and associated data treatment methods have also been developed for monitoring and analyzing metals, especially heavy metals in polluted soils, showing good performance for elements such as Ba, Co, Cr, Cu, Mn, Ni, Pb, V, and Zn[19-22]. Although the importance of the targeted applications leaves no room for doubt, the above demonstrations have not yet today led to large scale applications in real situation. The limited measurement precision and accuracy[23] that can be guaranteed by a LIBS instrument would represent a bottleneck issue for application of the technique, especially in the case of soil test. Indeed, the quantitative analysis capability of LIBS is still considered as its Achilles' heel[24].

In particular, for soil test, precision and accuracy of the measurements would be affected by a mediocre sample-to-sample repeatability of different measurements on i) samples of a given type of soil and ii) samples from different types of soil. Such repeatability is greatly influenced by the complex nature of laser-sample interaction, which depends upon both the laser characteristics and the sample material properties[25]. In a measurement, any uncontrolled change of the conditions of laser-sample interaction (laser pulse energy, laser pulse focusing, laser pulse space and/or temporal profile …) can lead to changes in the property of generated plasma (ablation rate, atomization yield, excitation temperature…), causing the so-called emission source noise[26]. Furthermore, inhomogeneity of a soil sample, even prepared in pellet after being ground into fine particles (of about 100 μm in size), can also induce changes in the plasma property and thus contribute to emission source noise[27], when different positions on the surface of a soil sample pellet are ablated by laser. In other frequent cases of analyzing different types of soil, the change of plasma property under the same experimental condition because of change of sample matrix, more specifically refers to matrix effect, which leads the emission intensity of a given element to change according to its compound speciation in the sample and the composition of the soil[28]. Although the matrix effect represents a general issue in LIBS[29,30], its influence in analysis of soil becomes much more pronounced because of the complex physical and chemical property and the associated wide range of different types of soil[31,32].

It is therefore crucial, for LIBS analysis of soils, to reduce and correct fluctuations of spectral intensity caused by the emission source noise and the matrix effect. Judicious sample preparation and correct use of internal reference may lead to significant improvement of the repeatability of a LIBS instrument, thus the precision and the accuracy of the measurements[33], although such preparation is not always possible nor efficient in case of soil analysis because of the abovementioned complexity of soil and practical constraints related to *in situ* and/or online measurements. Post-acquisition data treatment remains often the only efficient way for analytical performance improvement. Multivariate regression based on chemometry, principally partial least-squares regression (PLSR) and neuronal networks analysis (NNA), has been demonstrated being able to provide robust calibration models for soil samples, with furthermore a reduced dependence of such models on the specific soil chemical and physical properties[34-38].

In this work, we use artificial intelligence approach to further significantly improve the data processing of LIBS spectrum of soil with a particular concern in the establishment of a soil-independent calibration model able to efficiently take into account samples from different types of soil, and in the same time to significantly reduce the influence of emission source noise. One of the key points is to introduce the concept of generalized spectrum which includes usual spectral intensities and additional parameters containing the information about the sample (type of soil, sample preparation method…). A machine learning algorithm which offers a flexible and versatile framework to deal with heterogeneous data types has been used to develop multivariate calibration models. In the following, we will first present the raw experiment data and the analytical performances with a univariate calibration model.

The principle of the developed multivariate data processing method is then presented in detail. The analytical performance with the resulted multivariate calibration model are described and compared with those of the univariate model. By such comparison, we emphasize the satisfactory and impressive reduction of emission source noise and matrix effect allowed by the developed machine learning-based multivariate calibration model, before we deliver the conclusion of the paper.

# Experimental Data and Analytical Performances with Univariate Calibration Model

**Laboratory-prepared reference samples.** In the experiment, LIBS measurement was performed with 4 types of soils: NIST 2710 (called N1 in this paper), NIST 2587 (N2), collected 1 (U1) and collected 2 (U2). The corresponding powders were first spiked using a standard reference solution of silver in order to prepare a set of laboratory-prepared reference samples for each soil with 7 different Ag (as the analyte) concentrations in the range from 20 to 840 ppm weight. Pellets were then prepared with doped powders for every Ag concentration ($Co_{ti}$) of each of the 4 soil types ($t$). Table 1 shows the concentrations in ppm weight of the pellets and their role in the construction and validation of the calibration model. More details about sample preparation are provided in the section "Methods".

| Soil type $t$ | Calibration sample set (5 concentrations in ppm weight each soil, $Co_{ti}$) | Validation sample set (2 concentrations in ppm weight each soil, $Co'_{ti}$) |
|---|---|---|
| | ($i$) 1, 3, 4, 5, 7 | 2, 6 |
| NIST 2710 (N1) *initially containing 40 ppm weight of Ag* | 60, 140, 240, 440, 840 | 90, 640 |
| NIST 2587 (N2) | 20, 100, 200, 400, 800 | 50, 600 |
| Collected 1 (U1) | | |
| Collected 2 (U2) | | |

Table 1. Silver concentrations in ppm weight of the prepared sample pellets with their roles in the construction and validation of the univariate calibration model.

**Raw experimental data.** Six replicate ($j$) spectra were taken for each pellet. A spectrum was an accumulation of 200 laser shots distributed over 10 distinguished sites ablated each by 20 consequent laser pulses. An individual spectrum can thus be notated by $\vec{I_{ij}^t}$ (for the $j^{th}$ replicate measurement on the sample with analyte concentration $Co_{ti}$ prepared with the soil type $t$). A typical replicate-averaged spectrum is presented in Fig. 1, showing in particular the emission line chosen for Ag emission intensity measurement, the Ag I 328.1 nm line.

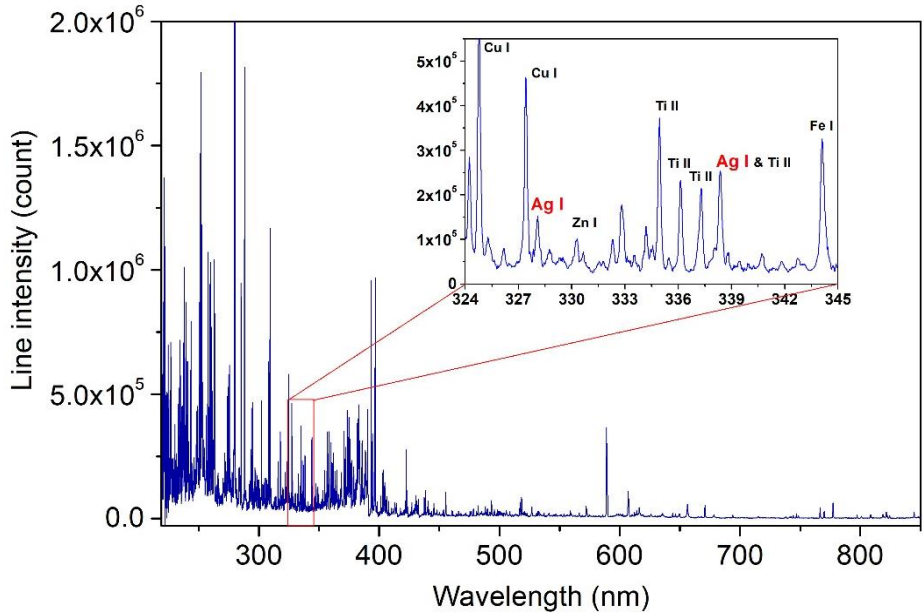

**Figure 1.** Typical replicate-averaged spectrum of soil sample. In the inset, the detailed spectrum around the Ag I 328.1 nm line is shown. Sample used to obtain the spectrum: $t$ =N1, initially containing the following elements: Cu (3420 ppm), Zn (4180 ppm), Ti (3110 ppm), Fe (43200 ppm), and Ag (40 ppm), 400 ppm of Ag was additionally spiked into the sample ($Co_{ti}$=440 ppm).

**Quantitative analysis performances with univariate calibration**. Calibration curves based on univariate regression are constructed by representing the intensity of Ag I 328.1 nm line, $\vec{I_{ij}^t}$(Ag I 328.1 nm), as a function of Ag concentration for the calibration sample set. As we can see in Fig. 1, this line is enough intense and seems free of interference with other lines, its intensity is still not particularly high to avoid significant self-absorption to occur. Line intensity as a function of Ag concentration and the resulted soil-specific calibration curve for each of the 4 analyzed soils are shown in Fig. 2, the error bars in the figure are standard deviations ($\pm\sigma_{I_i^t}$) of the intensities calculated for the 6 replicate measurements performed for each sample pellet. We can see in Fig. 2 for all the soils, large error bars on each measured line intensity, which is due to the emission source noise as we discussed above. The same noise leads to reduced values of determination coefficient $r^2$ with respect to the unit. In addition, the slopes of the calibration curves are significantly different, showing an obvious matrix effect in LIBS analysis of the 4 soils. By merging the intensity data from the calibration sample set of the 4 types of soil, a soil-independent calibration curve can be established. For this purpose, the intensity data from the 4 soils are plotted as a function of Ag concentration in a same figure as shown in Fig. 3. We can see in this figure that a large dispersion of emission intensities for a given Ag concentration due to the matrix effect leads to a much reduced $r^2$ value of the univariate calibration curve.

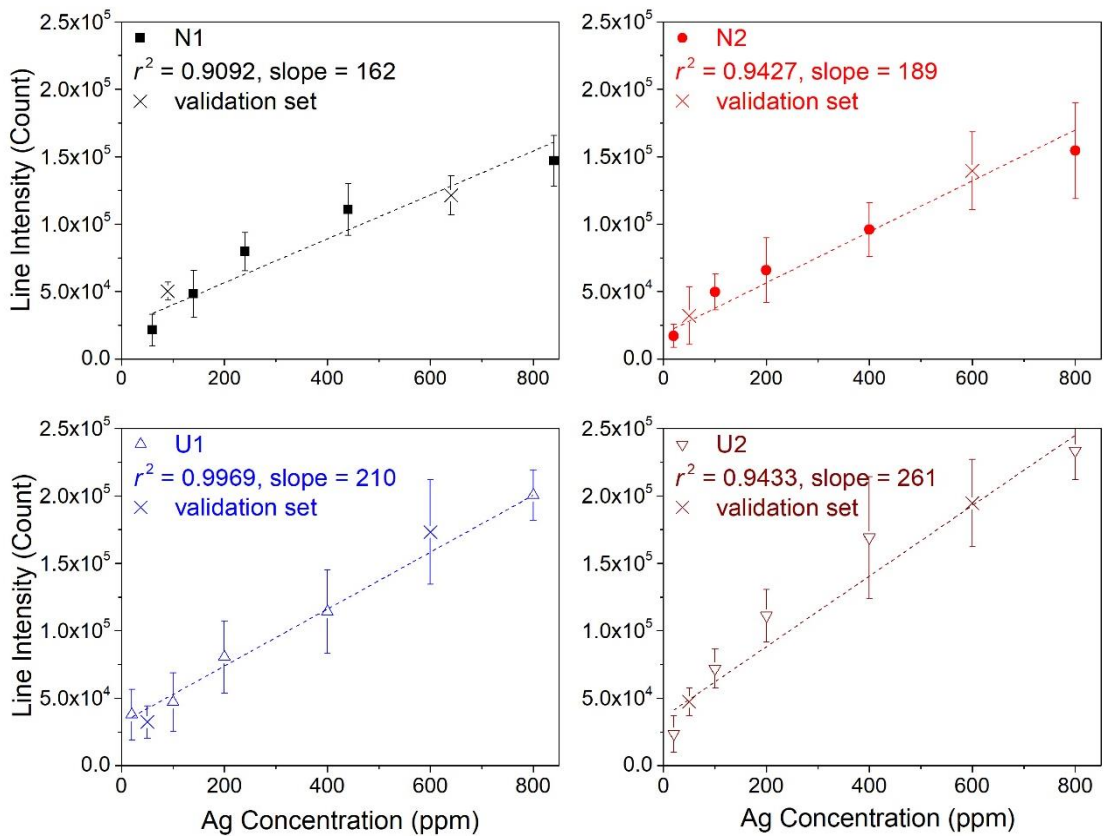

Figure 2. Intensity of Ag I 328.1 nm line of the calibration sample set as a function of Ag concentration and soil-specific univariate calibration curve (dashed lines in the figures) of Ag with this line respectively for the 4 analyzed soils. Line intensities from the validation set are represented by crosses, they do not participate in the construction of the calibration models. The error bars are calculated for each line intensity with the standard deviation among the 6 replicate measurements ($\pm \sigma_{I_i^t}$).

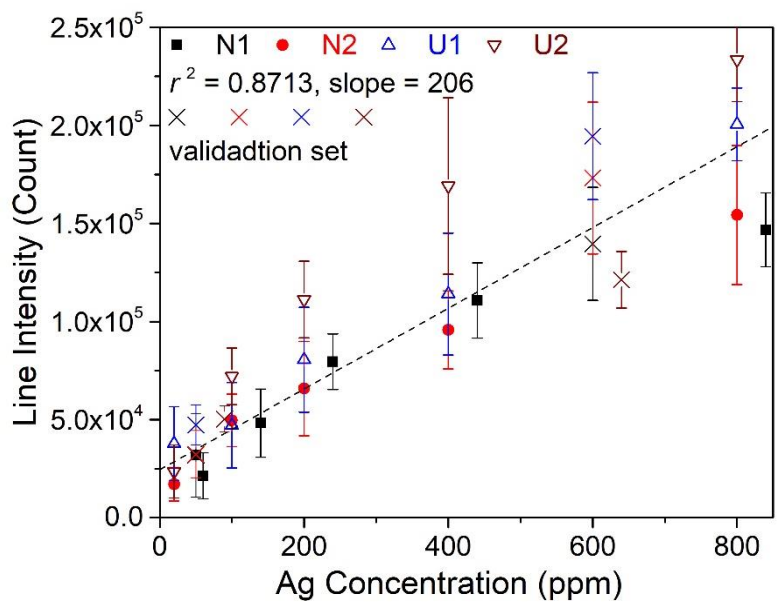

**Figure 3.** Similar presentation of the experimental data as in Fig. 2, but with line intensities from all the 4 soils merged in a same figure and a soil-independent univariate calibration curve (dashed lines in the figure).

Line intensities from the validation sample set in Table 1 are then used to evaluate the accuracy and the precision of prediction using the established calibration models. These intensities are represented by crosses in Fig. 2 and Fig. 3. Table 2 sums up the figures of merit[35,39] of quantitative analysis performance using the univariate model with both the soil-specific and the soil-independent calibration curves, where $REC(\%)$ is average relative error of calibration, $REP(\%)$ average relative error of prediction, $RSD(\%)$ relative standard deviation of the predicted concentrations, and $LOD$(ppm) limit of detection. The definitions of the above quantities are given in the section "Methods".

We can see in Table 2 that the soil-specific calibration curves have fair $r^2$ values, while their slopes are significantly different, as also shown in Fig. 2, indicating significant influences of both emission source noise and matrix effect. The accuracies of calibration and prediction, indicated respectively by $REC$ (mean value 18.28%) and $REP$ (mean value 37.07%) of the soil-specific calibration curves are not satisfactory. This is due to a limited measurement repeatability from one sample to another. On the other hand, a limited repeatability of replicate measurements leads to an unsatisfactory prediction precision (mean value of $RSD = 42.35\%$), as well as a quite high $LOD$ (mean value of $LOD = 24.23$ ppm), compared to standard LIBS measurement performances for solid samples.

| Calibration type | Soil | Calibration model | | | | Validation | |
|---|---|---|---|---|---|---|---|
| | | $r^2$ | Slope | $REC$(%) | $LOD$(ppm) | $REP$(%) | $RSD$(%) |
| Soil-specific | N1 | 0.9092 | 162 | 26.35 | 27.57 | 30.75 | 34.95 |
| | N2 | 0.9427 | 189 | 15.15 | 18.47 | 23.99 | 82.29 |
| | U1 | 0.9969 | 210 | 16.71 | 31.90 | 54.43 | 25.58 |
| | U2 | 0.9433 | 261 | 14.91 | 18.96 | 39.09 | 26.56 |
| | Mean | 0.9480 | 206 | 18.28 | 24.23 | 37.07 | 42.35 |
| Soil-independent | N1 | 0.8713 | 206 | 32.15 | 23.83 | 15.42 | 27.59 |
| | N2 | | | | | 30.33 | 55.86 |
| | U1 | | | | | 36.50 | 24.37 |
| | U2 | | | | | 42.54 | 50.32 |
| | All | | | | | 31.20 | 51.20 |

Table 2. Figures of merit of quantitative analysis performance of the univariate calibration model with the both soil-specific and soil-independent calibration curves.

When the soil-independent calibration curve is assessed, we can find a degraded $r^2$ value, indicating a significant influence of matrix effect. The extracted calibration curve slope logically corresponds to the average value of the slopes of the 4 soil-specific calibration curves. We can also see that the mentioned matrix effect also degrades the accuracy of calibration ($REC$) due to a larger dispersion of the line intensities participating in the construction of the soil-independent calibration curve. The degraded calibration accuracy becomes comparable to the prediction accuracy ($REP$) when all the soil types are considered, since both of them are directly influenced by the matrix effect. Due to the same influence, the all-soil-type prediction precision ($RSD$) is decreased compared to the mean value of the soil-specific calibration curves. At the same time, the limit of detection ($LOD$) does not record significant change with respect to the soil-specific calibration models, since it is more sensitive to the fluctuation of replicate measurements due emission source noise.

# Analytical Performances with Multivariate Calibration model

**Principle of the developed multivariate calibration model.** From the above results with univariate calibration models, we can see that the analytical performance is affected by both matrix effect and emission source noise. The developed multivariate model is therefore designed to deal with different types of soil, and in the same time, such model should efficiently reduce the dispersion of analytical results due to any change and fluctuation of experimental condition. The idea is to explicitly include information about soil type in the input variable of the calibration model. More specifically, for a given reference sample with known analyte concentration $Co_{ti}$ prepared from the soil type $t$, the $j^{th}$ replicate LIBS measurement generates a spectrum which can be presented as a vector in the form of $\overrightarrow{I_{ij}^t} = (I_{ij1}^t, I_{ij2}^t, \ldots I_{ijk}^t, \ldots I_{ijM_1}^t)$, where $M_1$ is the dimension of the spectrum (pixel number of a raw spectrum or the number of contained intensities in a pretreated spectrum). Such physical spectrum can be concatenated with an ensemble of $M_2$ variables, $(Ma_1^t, Ma_2^t, \ldots Ma_{M_2}^t)$, representing the properties of the sample. The result is a generalized spectrum with $M_1 + M_2$ generalized intensities,

$$\overrightarrow{I_{ij}^{t,Genral}} = (I_{ij1}^t, I_{ij2}^t, \ldots I_{ijM_1}^t, Ma_1^t, Ma_2^t, \ldots Ma_{M_2}^t). \tag{1}$$

Such spectrum is considered in the method as a vector in a hyperspace of $M_1 + M_2$ dimensions. A generalized module, $\left|\overrightarrow{I_{ij}^{t,General}}\right|_{General}$, can thus be attributed to it for formally representing the concentration of analyte (silver for instance) in the corresponding sample. Obviously such module cannot be calculated using a simple mathematical function. The physical correlation between the generalized spectrum and the concentration of the analyte can only be expressed as a mathematical relation of mapping:

$$f: \mathbb{R}_+^{M_1+M_2} \to \mathbb{R}_+, \ \overrightarrow{I_{ij}^{t,General}} \mapsto Co_{ti} = \left|\overrightarrow{I_{ij}^{t,General}}\right|_{General}. \tag{2}$$

In our experiment, machine learning is used through a training process, to establish mapping between the collection of generalized spectra and the ensemble of element concentrations of the corresponding reference samples. The result of such training process leads to a calibration model which is able to predict the concentration of the analyte in a validation sample when its generalized spectrum is used as the input of the model. The physical basis of the existence of mapping between generalized spectra and elemental concentrations is the interaction between the different species in a laser-induced plasma, which leads to the correlation of the concentration of a specific element contained in the plasma to the whole plasma emission spectrum.

**Implementation of the method: flowchart of training and validation of the calibration model**. Figure 4 presents the flowchart of the developed multivariate calibration method. Several steps can be distinguished in a successive way.

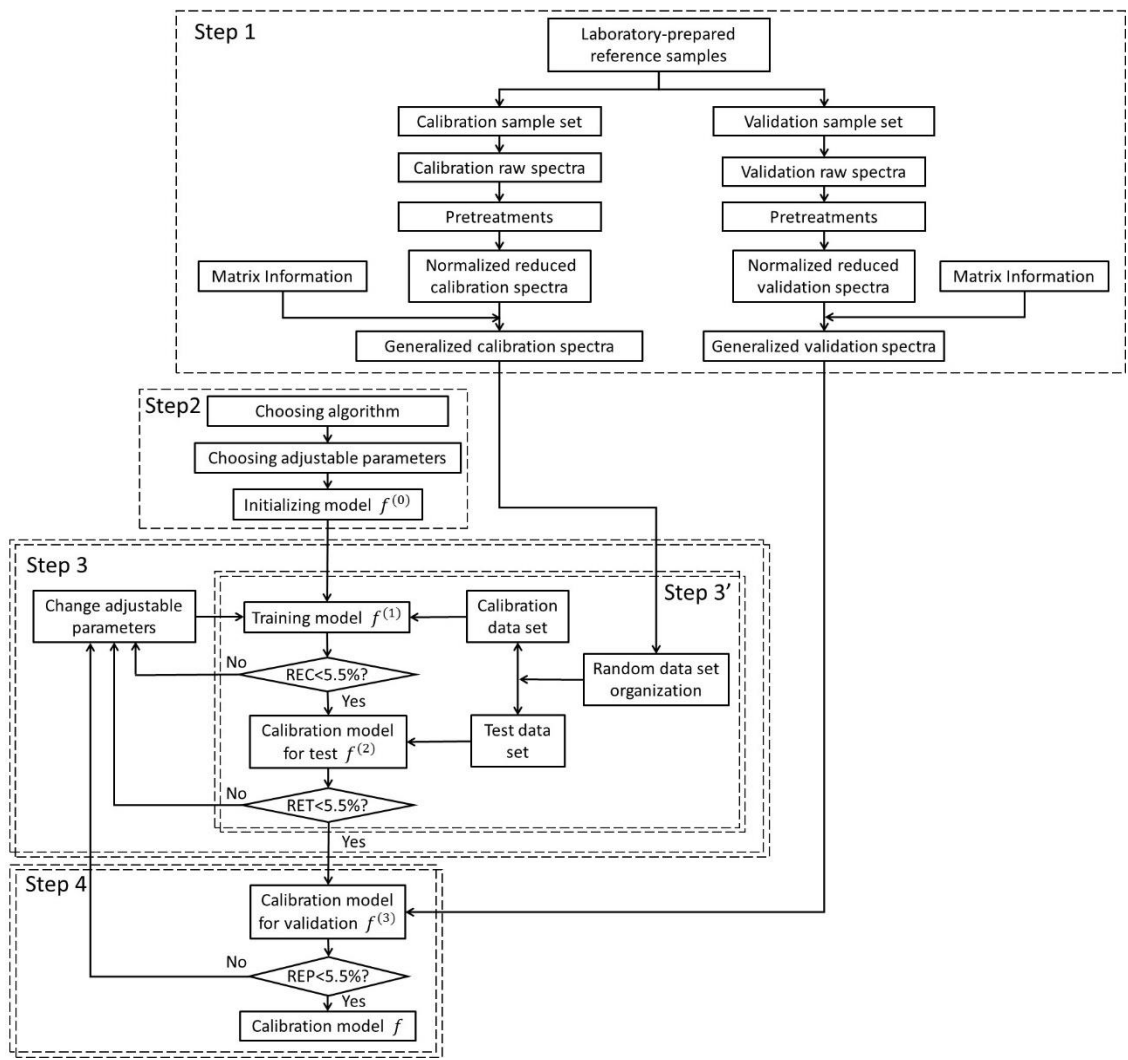

**Figure 4.** Flowchart for the built-up of the multivariate calibration model. The steps contained in double dashed line rectangles are repeated within a conditional loop.

***Step 1. Data set organization, pretreatment and formatting.*** The experimental data are organized in this step in the way shown in Table 3, where we can see that for each type soil, 6 pellets with different analyte concentrations are selected as calibration sample set ($i \in \{1,2,3,5,6,7\}$), and the rest one ($i \in \{4\}$) as validation sample set. In order to have a clear vision of the structure of the experimental data, they are presented within a rectangular parallelepiped as shown in Fig. 5. An individual raw spectrum, $\vec{I_{ij}^t} = (I_{ij1}^t, I_{ij2}^t, \ldots I_{ijk}^t, \ldots I_{ijM_0}^t)$, is represented in the rectangular parallelepiped by a cube with a set of given values of $(t, i, j)$, here the index $k$ is used to indicate a pixel in the spectrum: $1 \leq k \leq M_0$, $M_0 = 21915$ is the pixel number of a raw spectra, which physically corresponds to the spectral range of the used spectrometer, $220 \text{ nm} \leq \lambda \leq 850 \text{ nm}$.

| Soil type $t$ | Calibration sample set (6 concentrations in ppm weight each soil, $Co_{ti}$) | Validation sample set (1 concentration in ppm weight each soil, $Co'_{ti}$) |
|---|---|---|
| | ($i$) 1, 2, 3, 5, 6, 7 | 4 |
| NIST 2710 (N1) | 60, 90, 140, 440, 640, 840 | 240 |
| NIST 2587 (N2) | 20, 50, 100, 400, 600, 800 | 200 |
| Collected 1 (U1) | | |
| Collected 2 (U2) | | |

**Table 3.** Organization of the experimental data for the built-up of the multivariate calibration model.

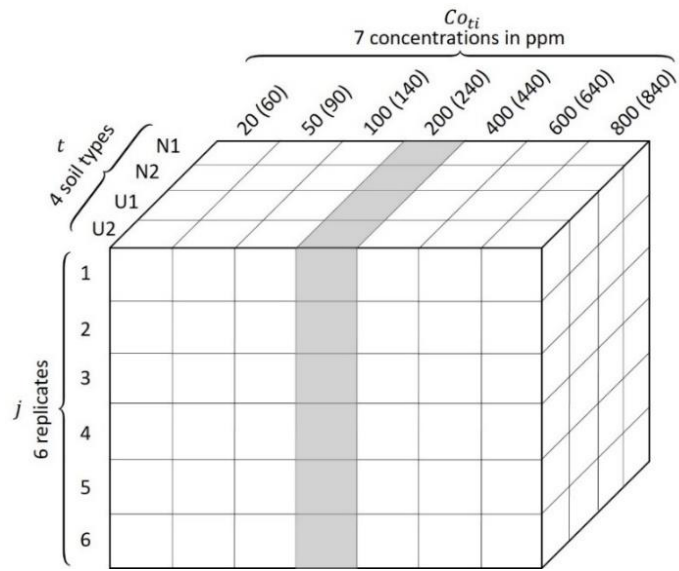

**Figure 5.** Structure of the experimental data with 4 soil types ($t$), 7 analyte concentrations ($Co_{ti}$) for each soil type and 6 replicate LIBS measurements ($j$) for a sample pellet of given soil type and analyte concentration. The samples with 200 (240) ppm analyte concentration are chosen as the validation sample set, the rest as the calibration sample set.

Pretreatment is performed on the raw spectra, which consists in i) normalization and ii) feature selection. Normalization, applied to all the raw spectra of the laboratory-prepared reference samples, is a simple operation which transforms the intensity rang of each pixel of all the raw spectrum into the interval between 0 and 1:

$$I_{ijk}^{t,norm} = \frac{I_{ijk}^t - I_k^{min}}{I_k^{max} - I_k^{min}} \quad \text{for } 1 \leq k \leq M_0, \tag{3}$$

where $i \in \{1,2,3,4,5,6,7\}$, $I_k^{min}$ and $I_k^{max}$ are respectively the minimum and the maximum of the pixel $k$ among the same pixels of all the individual spectra ($4 \times 7 \times 6 = 168$ spectra). Such normalization reduces the contrast among the pixel intensities of a raw spectrum, which can exceed one order of magnitude for a large part of the pixels as shown in Fig. 1. Since one could expect smaller variations among the intensities of the different individual spectra for a given pixel, unless a physical reason, analyte concentration variation for example, make them to change in a correlated way. After the normalization, all the pixels, whatever their initial physical intensities, should contribute in a more statistically equivalent way, to characterize an individual spectrum with respect to the others.

Feature selection is performed by applying the SelectKBest algorithm[40] to the normalized spectra of the calibration sample set. The principle consists in selecting and keeping in an individual spectrum for the further processing, pixel intensities with high enough correlation with the series of analyte concentrations of the calibration sample set. Such correlation is calculated in the algorithm with a score function $Score(k_0)$:

$$Score(k_0) = \mathcal{D} \frac{Corr(k_0)^2}{1 - Corr(k_0)^2}, \quad \text{for } 1 \leq k_0 \leq M_0, \tag{4}$$

$$Corr(k_0) = \frac{Cov(\{I_{ijk_0}^{t,norm}\}, \{Co_{ti}\})}{\sqrt{Var(I_{ijk_0}^{t,norm}) Var(Co_{ti})}}, \tag{5}$$

$$Cov(\{I_{ijk_0}^{t,norm}\}, \{Co_{ti}\}) = \frac{1}{\mathcal{D}} \sum_{i \text{ in } s} \sum_{t=1}^{4} \sum_{j=1}^{6} [(I_{ijk_0}^{t,norm} - \overline{I_{k_0}^{norm}})(Co_{ti} - \overline{Co})], \tag{6}$$

$$Var(I_{ijk_0}^{t,norm}) = \frac{1}{\mathcal{D}}\sum_{i \text{ in } S}\sum_{t=1}^{4}\sum_{j=1}^{6}(I_{ijk_0}^{t,norm} - \overline{I_{k_0}^{norm}})^2, \quad (7)$$

$$Var(Co_i) = \frac{1}{\mathcal{D}}\sum_{i \text{ in } S}\sum_{t=1}^{4}\sum_{j=1}^{6}(Co_{ti} - \overline{Co})^2, \quad (8)$$

where $S = \{1,2,3,5,6,7\}$, $\mathcal{D} = 6 \times 4 \times 6 = 144$ is the number of the individual spectra in the calibration sample set, $\overline{I_{k_0}^{norm}}$ stands the mean value of normalized intensity of the pixel $k_0$ (hence the corresponding wavelength) with respect to the measurement replicates, the soil types and the prepared concentrations of the calibration sample set; and $\overline{Co}$ refers to the mean value of the prepared concentrations of the calibration sample set. In the case of model training with a given type of soil, the above sums with respect to $t$ reduce to a single corresponding term. The threshold value applied to $Score(k_0)$ for feature selection takes into account the number of individual spectra included in the calibration sample set, for instance $\mathcal{D} = 6 \times 4 \times 6 = 144$. In this work, 150 pixels were selected over the initial 21915 ones, so that the reduced normalized spectrum have a dimension of $M_1 = 150$. Such dimension is comparable to the total number of spectra used in the calibration sample set, avoiding thus overfitting.

The spectrum of selected features is shown in Fig. 6. We can see that pixels (or equivalently wavelengths) receiving high scores are concentrated in the spectral range from 327 nm to 339 nm as shown in Fig. 6(b), and correspondingly in Fig. 6(a). We can identify 2 neutral silver lines: Ag I 328.1 nm with a NIST relative intensity[41] of 55000 and Ag I 338.3 nm with a smaller NIST relative intensity of 28000. The experimental spectrum shows however $I$(at 338.3 nm) > $I$(at 328.1 nm). A detailed inspection in the NIST Atomic Spectra Database[41] shows the presence of a relatively intense titanium ionic line, Ti II 338.4 nm line, with a NIST relative intensity of 7100. Since titanium represents an important trace element in soil, this line can therefore significantly interfere with the Ag I 338.3 nm line. This is why the pixels in the Ag I 328.1 nm line receive the highest scores, while a part of the pixels corresponding to the Ag I 338.3 nm line receive lower scores, and those pixels are all situated in the low frequency side of the intensity peak around 338.3 nm. A second zone where high score features are found extends from 750 nm to 850 nm, where we can remark the correspondence between the selected features and the lines emitted by oxygen and nitrogen atoms, which are mainly contributed by the ambient gas. Correlation between the elements from the ambient gas (especially O and N) and an element to be detected in the sample has been studied in our previous work[33]. Clear physical interpretation of the selected spectral features demonstrates the significance of the SelectKBest algorithm.

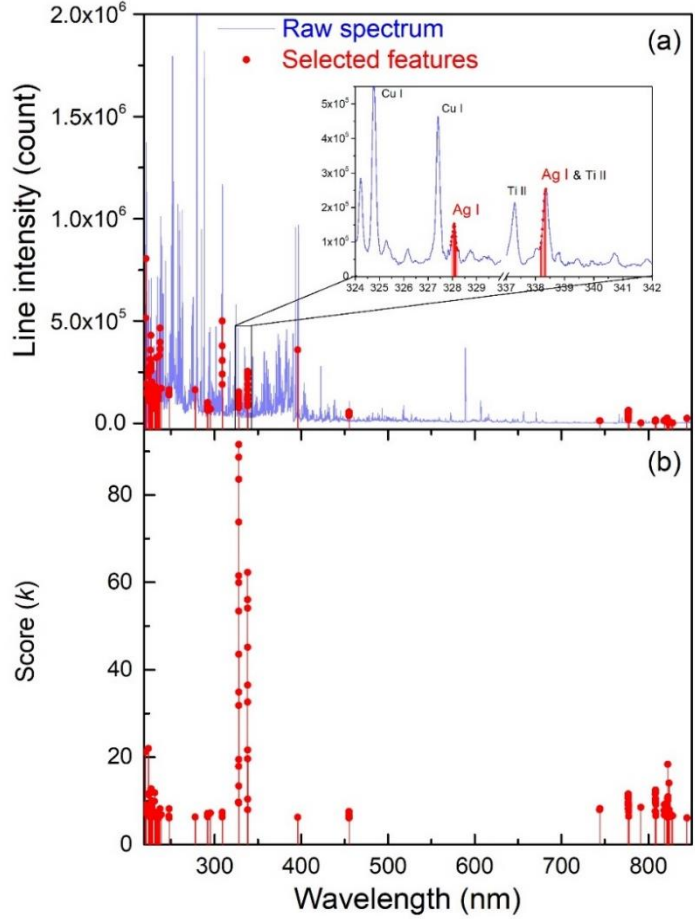

**Figure 6.** (a) Spectrum of the selected features (in red) with in the inset, those corresponding to the 2 Ag I lines, the raw spectrum (in light blue) is also shown for comparison; (b) Spectrum of the SelectKBest scores.

In our experiment, the type of soil is the only significant information which distinguishes the 4 soils (the same preparation procedure), it is thus concatenated with the normalized and reduced spectrum to form generalized spectrum: $\overrightarrow{I_{ij}^{t,general}} = \left(I_{ij1}^{t,norm}, I_{ij2}^{t,norm}, \ldots I_{ij150}^{t,norm}, Ma_1^t\right)$. Numerical values of $Ma_1^t = 1, 2, 3$ and 4 are arbitrarily chosen for representing the 4 soil types, N1, N2, U1 and U2 respectively.

***Step 2. Model initialization.*** Back-propagation artificial neural networks (BPNN)[42] is chosen in this work to provide the algorithm which maps generalized spectra and corresponding analyte concentrations. The number of hidden layers $n\_layers$ and the number of nodes in a layer $n\_nodes$ are selected as the externally adjustable parameters to optimize the performance of the model. The model starts with its default state denoted by $f^{(0)}$.

***Step 3. Model built-up loop: training of the algorithm and optimization of its externally adjustable parameters.*** This is the central body of the model construction process which comprises an internal and an external loop as shown in Fig. 4.

The internal loop (Step 3' in Fig. 4) is devoted to train the algorithm in such way for an input individual generalized spectrum $\overrightarrow{I_{lJ}^{t,general}}$, the resulted generalized module becomes as close as possible to the targeted analyte concentration $Co_{ti}$. In considering the statistical equivalence and experimental fluctuation among the replicate measurements ($j$ is a dummy index) and the matrix effect due to different soils, and in order to fulfill the requirements for the model to tackle both the experimental fluctuation and the matrix effect, the training process takes place in the following way:

i) Randomly permuting among $j$ of all the data columns of given $t$ and $Co_{ti}$, in order to randomly and independently fix the arrangement of all the 24 columns of replicate spectra as shown in Fig. 7a, with the arrangements visible for all the columns in the surface of $t = U2$ and $Co_{ti} = 800(840)$ ppm of the date cube;

ii) For one of the data configurations (in total $(6!)^{24}$ possible and statistically equivalent ones) generated in the above way, performing a dynamic cross validation training process of 6 iterations. In each of these iterations, successively one layer of the data, for example the top layer, then the second, then the third…, up to the bottom one, is considered as test data set, while the rest as calibration data set as shown in Fig. 7b. In such iteration, the algorithm corresponding to a training model, $f^{(1)}$, is trained, with the calibration data set, in order for the output generalized modules of the individual generalized spectra to be as close as possible to the corresponding target silver concentrations. These iterations generate 6 different BPNNs.

iii) In the end of the above 6-fold iterative training and cross validation process,

another randomly and independently arranged data configuration is generated for a new 6-fold iterative cross validation training of the algorithm. In the experiment, we fixed the considered number of randomly and independently arranged data configurations to 10, because a larger number of data configurations would not significantly enrich useful information that we can extract from the given ensemble of raw experimental spectra. In the end of the 10-data-configuration training, 60 different BPNNs are generated.

iv) The average relative error of calibration ($REC$) is calculated. If the value is larger than the fixed threshold, the process goes back to the training step of $f^{(1)}$. Otherwise a calibration model for test, $f^{(2)}$, is generated.

v) $f^{(2)}$ is then tested by the test data set in a similar way as the above training process. The average relative error of test ($RET$) is calculated.

vi) The resulted $REC$ and $RET$ are compared to the fixed threshold values. If they, or one of them, are larger than the threshold value(s), the process goes to the external loop. Otherwise a calibration model for validation, $f^{(3)}$, is generated.

In this experiment, the threshold values were fixed for 10-data-configuration resulted $REC$ and $RET$ both at 5.50%. This value was chosen to minimize the average relative error of prediction ($REP$) calculated in the validation process of the calibration model, $f$, by generalized validation spectra, which were not involved in the model training process. Numerical experiments were thus necessary to determine these thresholds, even though the values could be intrinsically smaller if only the model training process in the step 3 is concerned.

The detailed definitions of $REC$, $RET$ and $REP$ for the assessment of the multivariate model are given in the section "Methods".

Then only one type of soil is under consideration, $t$ takes a fixed valued among $N1, N2, U1, U2$.

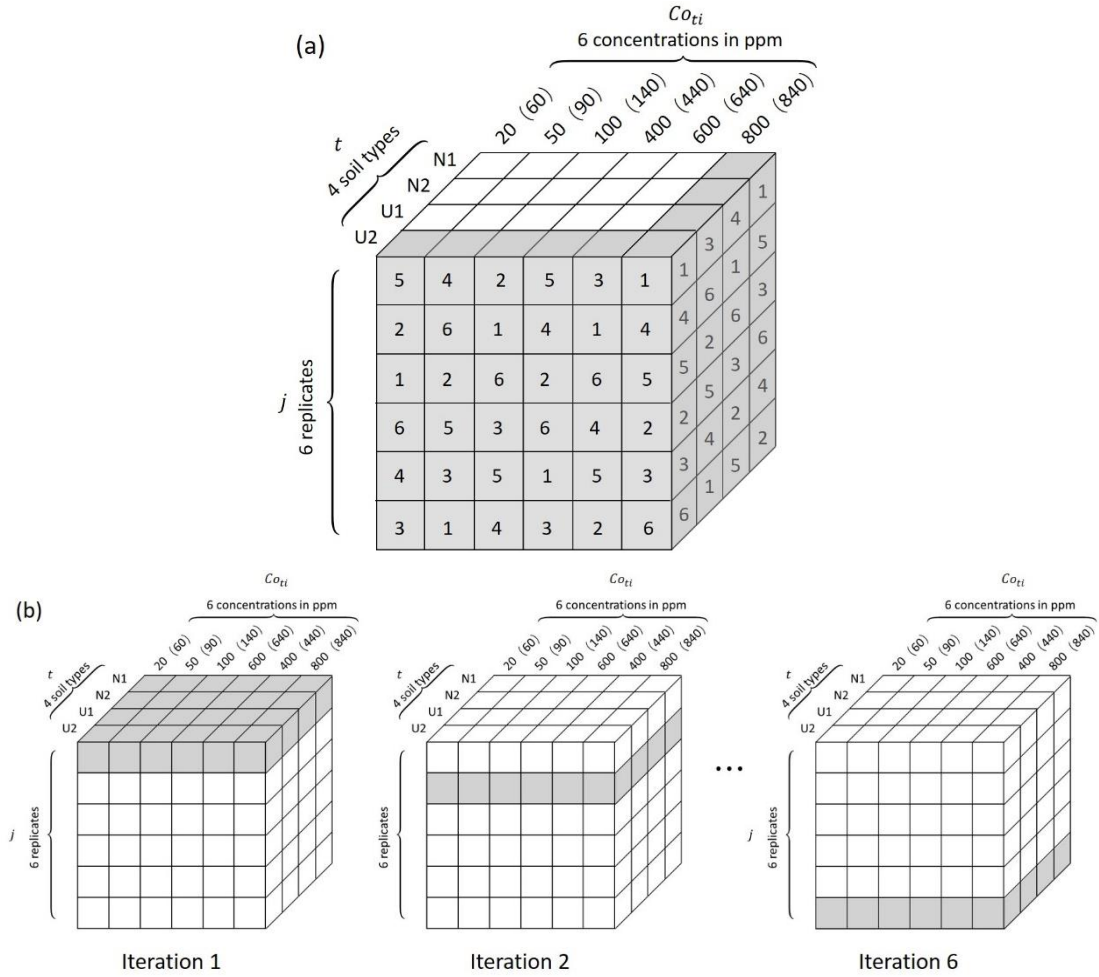

**Figure 7.** (a) A randomly and independently arranged data configuration among $(6!)^{24}$ possible and statistically equivalent ones; (b) For a given randomly and independently arranged data configuration, illustration of a 6-fold cross-validated training iteration, with the cubes in grey representing the test data set.

The external loop of this step is aimed to optimized the externally adjustable parameters of the algorithm, BPNN for instance. The used method is grid-search parameter tuning, which is known as an efficient method of optimization for constructing a calibration model. In this method, for given ranges of the selected adjustable parameters, the performance of the model is evaluated for all the possible combinations of the adjustable parameters in an exhaustive way. The combination generating the best performance is retained. In our experiment, the ranges of the 2 externally adjustable parameters, $n\_layers$ and $n\_nodes$, both positive integer, were respectively fixed being 1 to 2 and 3 to 8, 12 combinations were therefore evaluated.

When the values of $REC$ and $RET$ are simultaneously smaller than 5.50%,

the iteration in the external loop stops. A calibration model for validation $f^{(3)}$ is obtained as the output of the step 3.

***Step 4. Model validation with an independent validation sample set.*** The output model of the step 3, $f^{(3)}$, is validated in this step using generalized validation spectra obtained from the validation sample set which is not involved in the model training process. Average relative error of prediction ($REP$) and average relative standard deviation ($RSD$) are calculated for individual generalized validation spectra of the validation sample set to respectively evaluate the prediction accuracy and precision of the model. The resulted $REP$ is compared to a threshold value, which is fixed in our experiment at 5.50%. When $REP$ is larger than this threshold, the process returns back to the step 3, leading to further optimizations of the externally adjustable parameters and the algorithm by training and cross-validation. The process continues until $REP$ becomes smaller than the threshold value to generate the final calibration model, $f$. In our experiment, the final calibration model was generated with externally adjustable parameters of $n_{layers} = 1$ and $n_{nodes} = 5$.

**Results and discussions.** Soil-specific and soil-independent calibration curves are respectively shown in Fig. 8 and Fig. 9. We use here a similar presentation as in Fig. 2 and Fig. 3 to easy the comparison with the univariate models. And the parameters showing the analytical performances of the multivariate models are presented in Table 4.

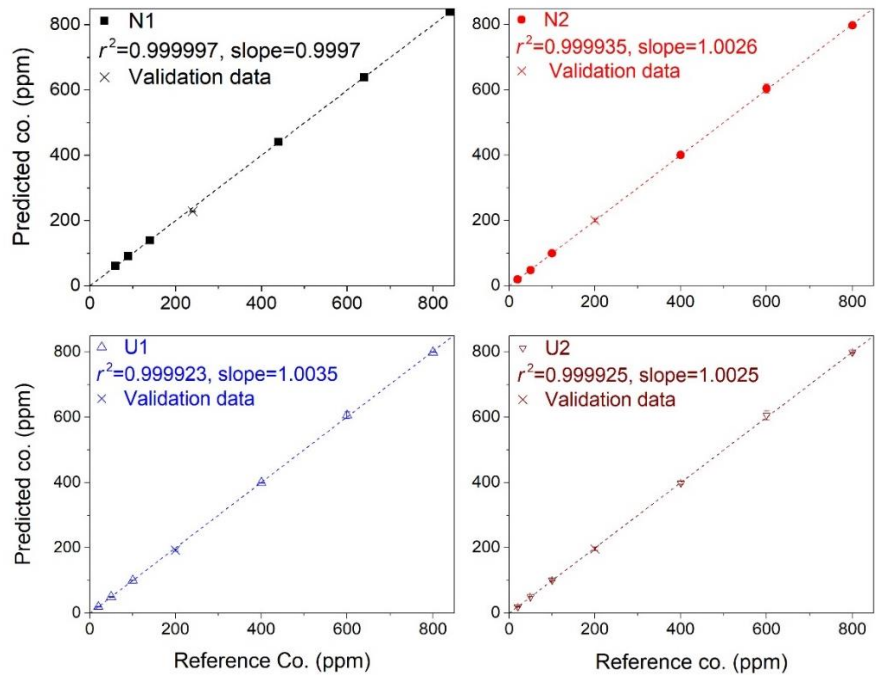

Figure 8. Model-predicted Ag concentrations as function of the prepared ones and soil-specific calibration curves for Ag concentration based on multivariate calibration models. Validation data are represented in the figure with crosses.

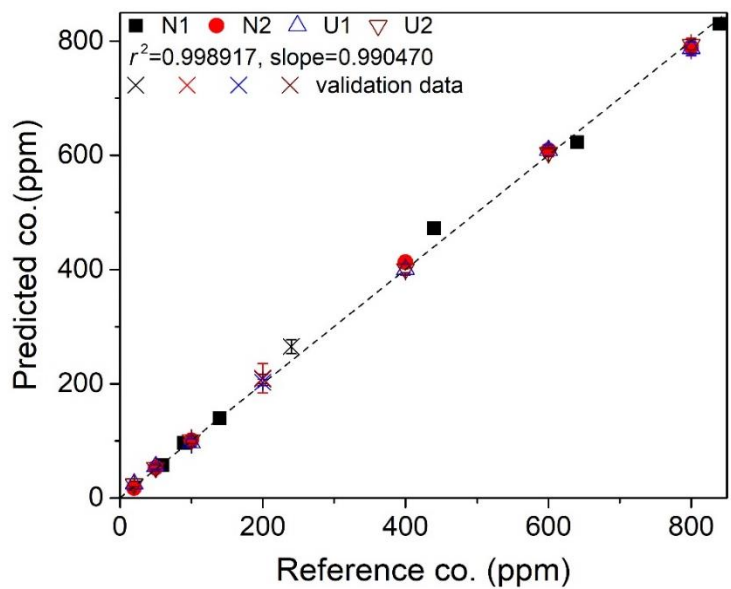

Figure 9. Model-predicted Ag concentrations as function of the prepared ones and soil-independent calibration curve for Ag concentration based on multivariate calibration model. Validation data are represented in the figure with crosses.

| Calibration type | Soil | Calibration model | | | | | Validation | |
|---|---|---|---|---|---|---|---|---|
| | | $r^2$ | Slope | REC(%) | RET(%) | LOD(ppm) | REP(%) | RSD(%) |
| Soil-specific | N1 | 0.999997 | 0.9997 | 0.084 | 0.176 | 1.158 | 4.97 | 0.89 |
| | N2 | 0.999935 | 1.0026 | 0.671 | 1.215 | 1.405 | 0.54 | 2.63 |
| | U1 | 0.999923 | 1.0035 | 0.447 | 0.991 | 0.710 | 3.37 | 1.88 |
| | U2 | 0.999925 | 1.0025 | 0.389 | 0.914 | 2.386 | 1.49 | 2.76 |
| | Mean | 0.999945 | 1.0021 | 0.398 | 0.824 | 1.415 | 2.59 | 2.04 |
| Soil-independent | N1 | 0.998917 | 0.9882 | 3.705 | 5.09 | 4.962 | 22.13 | 3.96 |
| | N2 | | | | | | 0.25 | 12.24 |
| | U1 | | | | | | 3.19 | 2.76 |
| | U2 | | | | | | 0.38 | 3.96 |
| | All | | | | | | 5.20 | 5.90 |

**Table 4.** Figures of merit of quantitative analysis performance of the mulitivariate calibration models with both the soil-specific and the soil-independent calibration curves.

We can see that the soil-specific calibration curves exhibit all a $r^2$ value very close to the unit. This means that the multivariate models efficiently reduce the experimental fluctuation from a reference sample to another. The fluctuation from a replicate measurement to another for a given sample is also significantly reduced, which leads to a very small error bar on each predicted concentration. A direct consequence of such reduced fluctuations is a significant improvement of $LODs$ from several tens ppm to around ppm. In coherence with the high $r^2$ values, the calibration accuracy is greatly improved and reaches now an impressive level of around 1% for $REC$ and $RET$. The prediction capacity of the soil-specific calibration models is clearly reinforced with order-of-magnitude reduction for both $REP$ and $RSD$ compared to those of the univariate model. Such performance clearly fulfills the requirements of precise and accurate quantitative analysis.

When the spectra from all the soils are used to build a calibration model, the soil-independent calibration curve is obtained as shown in Fig. 9. We can see a $r^2$ value very close to the unit as in the case of soil-specific multivariate calibration curves. This not only means an efficient improvement of the repeatability from a sample to another for a given type of soil, but more importantly shows the ability of the multivariate model to take into account the specific matrices between the different soils and to reduce the matrix effect. In fact, the data from the different types of soil can be fitted with a unique linear model with a determination coefficient $r^2$ very close to those of soil-specific calibration curves. The $LOD$ allowed by the soil-independent multivariate model remains quite low in the order of 5 ppm. A slight increase of this value with respect to those of soil-specific calibration curves would indicate a residual matrix effect. The same residual matrix effect should contribute to slightly reduce the calibration accuracy compared to the soil-specific calibration curves, as indicated by the values of $REC$ and $RET$ in the order of 5% for the soil-independent model. Compared to the univariate model, the performance of the calibration curve is greatly improved with a matrix effect reduced within an acceptable level.

Concerning the prediction capacity of the multivariate soil-independent model, great improvements can be observed with respect to the univariate model for accuracy as well as for precision, although degradations are observed compared to the multivariate soil-specific models. Such degradations should be related to the above mentioned residual matrix effect, which would lead to, sometimes, unexpected large values of $REP$ and $RSD$ for specific soils, which is the case for $N1$ (specified with an informative Ag initial concentration of 40 ppm weight from NIST) with a large $REP$ and $N2$ with a large $RSD$. Nevertheless, when the validation is extended to all the soils, the "average" prediction capacity exhibits an excellent level, as indicated by the values of corresponding $REP$ and $RSD$ in the range of 5 – 6%, which is order-of-magnitude improved compared to the univariate model. The degradations observed for the soil-independent model with respect to the soil-specific models seem suggesting possible improvements with a better correction of matrix effect, which might need an enlarged number of soil types used to train the multivariate model.

## Conclusions

In this work, a multivariate calibration model has been developed using spectra from LIBS measurements of laboratory reference samples prepared with 4 types of soil. The purpose is to strength the ability of quantitative analysis of LIBS technique by efficiently correcting fluctuation due to emission source noise and deviation due to matrix effect. In an application case as important as soil analysis, such fluctuation and deviation prevent univariate calibration model from being sufficient for precise and accurate quantitative analysis of contained trace elements. A multivariate calibration model has been therefore designed for taking into account the specificities of different soils and in the same time, efficiently reducing data dispersion due to experimental fluctuation. A key point is to introduce the concept of generalized spectrum, in which the information about sample matrix is explicitly included. A machine learning algorithm, BPNN, has been used to map a generalized spectrum to the corresponding analyte concentration. A training process, including data pretreatment, model initiation, model training loops and model validation, has been implemented within the framework of Python programing language. Especially in the data pretreatment, a feature selection algorithm reduces the dimension of a spectrum to a value compatible with the number of the raw spectra, avoiding thus overfitting, and in the same time extracts the most significant features for characterizing the spectrum.

The resulted multivariate model shows great improvements with respect to the univariate one. Fluctuation over the replicates is efficiently reduced, leading to very small error bars on the model-predicted concentrations. Correlated improvement of sample-to-sample repeatability for a given soil type further allows the soil-specific calibration curves exhibiting a $r^2$ value exceeding 0.9999, a calibration accuracy reaching 1% level, and a $LOD$ being down to the order of ppm. When being validated by independent samples, the prediction capacity of the soil-specific models presents high performance in terms of accuracy (mean $REP = 2.59\%$) as well as precision (mean $REP = 2.04\%$). When soil-independent calibration model is considered, the result of matrix effect correction is impressive with order-of-magnitude improvements with respect to the univariate model, for the calibration model as well as its validation with independent samples. Thereby, the accuracy and the precision of prediction are both improved into the range of 5 – 6%. Within the multivariate models, degradations can be observed for the soil-independent model when compared to the soil-specific models, the performance of the calibration curve ($r^2$, calibration accuracy and limit of detection) as well as the results of its validation (prediction accuracy and precision) are weakened, although staying satisfactory for quantitative analysis. This may indicate a residual matrix effect, a further reduction of which may need more soils with different matrices to be used in training of the multivariate model. Last but not least, once the model has been properly trained (it can take hours), the prediction for a spectrum from an unknown sample can be performed within a second using a standard personal computer.

Our work therefore demonstrates the pertinence and the advantage of applying machine learning to treat LIBS spectra of soils. In addition, the implementation schema of such approach is described in detail to easy any new applications of this method. The perspective to generalize the developed method to LIBS analysis of other materials, and furthermore to other spectroscopies is certainly worth to be mentioned here. Such generalization will indeed allow spectroscopic techniques, in the large sense of the term, benefiting from the wonderful progresses realized today and to be foreseen tomorrow in artificial intelligence. We believe therefore a breakthrough with the described approach for many applications, especially for online and/or *in situ* detection, monitoring and analysis, where experimental condition and wished sample preparation cannot be ideally controlled nor properly applied.

## Methods

**Soil samples and their preparation**. Four different soils were analyzed in the experiment. Two of them were standard reference materials (SRM) from National Institute of Standards and Technology (NIST): NIST 2710 (https://www-s.nist.gov/srmors/view detail.cfm?srm=2710a) and NIST 2587 (https://www-s.nist.gov/srmors/view detail. cfm?srm=2587), and respectively named as N1 and N2 in this work. The other 2 soil samples were collected from 2 different places near Lyon in France, one near a river (sand-like soil) and another in an agriculture field (yellow colored soil) with unknown elemental compositions and named as U1 and U2 in this work. The 2 NIST samples were provided in fine and uniform powder of particle size < 75 μm (200 mesh). The 2 collected samples were first dried, separated from small stones and organic materials, ground and then sequentially sieved through stainless steel sieves of 100, 200 and 400 mesh, assisted by an electromagnetic vibratory shaker, finally resulting particles with sizes of < 38 μm. In each type of soil powders, silver (Ag) as analyte was added in different concentrations, by mixing the soil powders with Ag solutions at different concentrations obtained by dilution with deionized water of an Ag standard solution (2 % nitric acid solution at an Ag concentration of 1000 mg/L from SPEX CertiPrep). Notice that the initial content of Ag in the collected soils was negligible (under the limit of detection of the experimental setup). For the 2 NIST samples, the N1 sample was specified with an informative initial silver concentration of 40 mg/kg (40 ppm weight). For the N2 sample, there was no specification of silver concentration. Doped powders were prepared in pellets of different soils and different Ag concentrations. For the preparation of a pellet, 0.2 g soil powder was pressed without binder under a pressure of 667 MPa (6.8 t/cm$^2$) for 5 min to form a pellet with a diameter of 13 mm.

**Experimental setup and measurement protocol**. The experimental setup used to produce the LIBS spectra has been described in detail elsewhere[33,43]. The following experimental parameters were used for the spectrum acquisition in this experiment: laser wavelength 1064 nm; laser pulse energy 60 mJ; diameter of the focused laser spot on the sample surface ~ 300 μm, estimated laser fluence on the sample surface 85 J/cm$^2$ and ablation under the atmospheric ambient. The emission from a zone around the symmetry axis of the plasma situated at a height of 1.3 mm from the sample surface was captured and coupled to an Echelle spectrometer (Andor Technology Mechelle 5000). The spectral range of the spectrometer was 220 nm – 850 nm, with a resolution power of $\lambda/\Delta\lambda \approx 5000$. The intensified CCD camera (ICCD) coupled to the spectrometer was triggered by laser pulse and set with a delay of 1 µs and a gate width of 2 µs. A gain of 60 (maximum 250) was applied of the intensifier of the ICCD for all the measurements. For each sample pellet of given Ag concentration of each type of soil, 6 replicate spectra were taken. Each spectrum was an accumulation of 200 laser shot distributed over 10 sites ablated each by 20 consequent laser pulses. Between 2 neighbor ablation sites, a translation stage displaced the pellet over a distance of 600 µm in order to avoid overlapping between the sites.

**Assessment of univariate calibration model**[35,39]. For a given type $t$ among the $T$ types of soil ($T = 4$ in this experiment), an ensemble of laboratory-prepared reference samples with different analyte concentrations are separated into a calibration sample set and a validation sample set:

$n$ ($n'$): number of reference samples with different concentrations $Co_{ti}$ ($Co'_{ti}$) prepared for the calibration (validation) sample set of soil type $t$, $1 \leq i \leq n$ ($1 \leq i \leq n'$), $n = 5$ ($n' = 2$) for the univariate model.

$J$: number of replicate measurements $j$, performed per calibration (or validation) sample, $1 \leq j \leq J$, $J = 6$ for the univariate model.

$I^t_{ij}$(Ag I 328.1 nm) ($I^{t'}_{ij}$(Ag I 328.1 nm)): experimentally recorded analyte emission intensity (called intensity for simplicity) from replicate measurement $j$ performed on a sample with concentration $Co_{ti}$ ($Co'_{ti}$).

$I^t_i$: mean value of experimental intensity corresponding to calibration sample $Co_{ti}$, $I^t_i = \frac{1}{J}\sum_{j=1}^{J} I^t_{ij}$(Ag I 328.1 nm).

$I_m$: mean experimental intensity for the calibration sample set of the $T$ soil types:

$I_m = \frac{1}{n \times T} \sum_{i \text{ in } S} \sum_{t=1}^{T} I^t_i$,

where $S = \{1,3,4,5,7\}$ referring to the calibration sample set.

$\hat{I}^t_i$: calculated intensity with the calibration model for a calibration sample with prepared concentration $Co_{ti}$.

$\widehat{Co}_{tij}$ ($\widehat{Co}'_{tij}$): predicted concentration with the calibration model (reverse calibration) for the experimental replicate intensity $I^t_{ij}$(Ag I 328.1 nm) ($I^{t'}_{ij}$(Ag I 328.1 nm)).

$\widehat{Co}_{ti}$ ($\widehat{Co}'_{ti}$): predicted concentration with the calibration model (reverse calibration) for

the experimental mean intensity $I_i^t$ ($I_i^{t'}$).

- Determination coefficient $r^2$ (the square of the correlation coefficient $r$), a usual criterion of the performance of a calibration model:

$$r^2 = 1 - \frac{SS_{res}}{SS_{tot}}, \qquad (9)$$

where $SS_{tot}$ is the sum of squares of the experimental intensities corrected by their mean value, $SS_{res}$ is the sum of squares of the residuals with respect to the calibration model:

$$SS_{tot} = \sum_{i \ in \ S} \sum_{t=1}^{T} (I_i^t - I_m)^2, \qquad (10)$$

$$SS_{res} = \sum_{i \ in \ S} \sum_{t=1}^{T} (I_i^t - \hat{I}_i^t)^2, \qquad (11)$$

where $S = \{1,3,4,5,7\}$ referring to the calibration sample set.

- Average relative error of calibration $REC(\%)$ for calibration accuracy evaluation:

$$REC(\%) = \frac{100}{n \times T} \sum_{i \ in \ S} \sum_{t=1}^{T} \left| \frac{\widehat{Co}_{ti} - Co_{ti}}{Co_{ti}} \right|, \qquad (12)$$

where $S = \{1,3,4,5,7\}$ referring to the calibration sample set.

- Average relative error of prediction $REP(\%)$ for prediction accuracy evaluation:

$$REP(\%) = \frac{100}{n' \times T} \sum_{i \ in \ S} \sum_{t=1}^{T} \left| \frac{\widehat{Co}'_{ti} - Co'_{ti}}{Co'_{ti}} \right|, \qquad (13)$$

where $S = \{2,6\}$ referring to the validation sample set.

- Relative standard deviation $RSD(\%)$ of the predicted concentrations for the validation sample set for prediction precision evaluation:

$$RSD(\%) = \frac{100}{n'} \sum_{i \ in \ S} \sqrt{\frac{1}{T \times J - 1} \sum_{t=1}^{T} \sum_{j=1}^{J} \left( \frac{\widehat{Co}'_{tij} - Co'_{ti}}{Co'_{ti}} \right)^2}, \qquad (14)$$

where $S = \{2,6\}$ referring to the validation sample set.

- Limit of detection $LOD$ (ppm), deduced by fitting the experimental intensity $I_{ij}^t$(Ag I 328.1 nm) versus prepared concentrations of the calibration sample set, $Co_{ti}$, by a straight line,

$$I_{ij}^t(\text{Ag I } 328.1 \text{ nm}) = a + b \times Co_{ti}, \qquad (15)$$

$$LOD(\text{ppm}) = \frac{3\sigma_a}{b}, \qquad (16)$$

where $\sigma_a$ is the standard deviation of $a$, such variation is due to the dispersion of $I_{ij}^t$(Ag I 328.1 nm). $LOD$ is thus determined by the sensibility of the technique (the slope $b$) and the repeatability and precision of intensity measurements among the different reference samples and different replicates for given samples (standard deviation of $a$, $\sigma_a$).

In the case of consideration of a specific soil type, the variable $t$ takes the corresponding given value and the concerned sum reduces to a specific term in the above definitions.

**Assessment of multivariate calibration model.** In the experiment, the multivariate calibration model, in its different training stages $f^{(q)}$, allows deducing a predicted analyte concentration $\widehat{Co}_{tij}^{(q)}$ when an individual generalized spectrum, $\overrightarrow{I_{ij}^{t,General}}$, of a sample with a laboratory-prepared analyte concentration $Co_{ti}$ (targeted concentration) is used as the input variable:

$$f^{(q)}: \mathbb{R}_+^{M_1+M_2} \to \mathbb{R}_+, \ \overrightarrow{I_{ij}^{t,General}} \mapsto \widehat{Co}_{tij}^{(q)} = \left|\overrightarrow{I_{ij}^{t,General}}\right|_{General}. \tag{17}$$

The following parameters are defined to assess the performance of the multivariate model:

$T = 4$: total number of soil type;

$n = 6$: number of different concentrations in the calibration sample set;

$n' = 1$: number of different concentrations in the validation sample set;

$J = 5$: number of replicates in the calibration data set;

$J' = 1$: number of replicates in the test data set;

$J'' = 6$: number of replicates in the validation sample set;

$O = 6$: number of iterations for a given randomly and independently arranged data configuration;

$P = 10$: number of randomly and independently arranged data configurations.

- Average relative error of calibration $REC(\%)$:

$$REC(\%) = \frac{100}{n \times T} \sum_{i \ in \ S} \sum_{t=1}^{T} \left|\frac{\widehat{Co}_{ti}^{(1)} - Co_{ti}}{Co_{ti}}\right|, \tag{18}$$

$$\widehat{Co}_{ti}^{(1)} = \frac{1}{P \times O \times J} \sum_{p=1}^{P} \sum_{o=1}^{O} \sum_{j=1}^{J} \left[\widehat{Co}_{tij}^{(1)}\right]_{(o,p)}, \tag{19}$$

where $S = \{1,2,3,5,6,7\}$ referring to the calibration sample set, $\left[\widehat{Co}_{tij}^{(1)}\right]_{(o,p)}$ is the predicted concentration corresponding to the targeted concentration $Co_{ti}$ by $f^{(1)}$ in a given iteration for a given randomly and independently arranged data configuration $(o,p)$:

$$f^{(1)}: \left[\overrightarrow{I_{ij}^{t,General}}\right]_{(o,p)} \mapsto \left[\widehat{Co}_{tij}^{(1)}\right]_{(o,p)}. \tag{20}$$

$\widehat{Co}_{ti}^{(1)}$ is the mean predicted concentration by $f^{(1)}$ with respect to the laboratory-prepared reference concentration $Co_{ti}$.

- Average relative error of test $RET(\%)$:

$$RET(\%) = \frac{100}{n \times T} \sum_{i \ in \ S} \sum_{t=1}^{T} \left| \frac{\widehat{Co}_{ti}^{(2)} - Co_{ti}}{Co_{ti}} \right|, \tag{21}$$

$$\widehat{Co}_{ti}^{(2)} = \frac{1}{P \times O \times J'} \sum_{p=1}^{P} \sum_{o=1}^{O} \sum_{j=1}^{J'} \left[ \widehat{Co}_{tij}^{(2)} \right]_{(o,p)}, \tag{22}$$

where $S = \{1,2,3,5,6,7\}$ referring to the calibration sample set, $\left[ \widehat{Co}_{tij}^{(2)} \right]_{(o,p)}$ is the predicted concentration corresponding to the targeted concentration $Co_{ti}$ by $f^{(2)}$ in a given iteration for a given randomly and independently arranged data configuration $(o, p)$:

$$f^{(2)}: \left[ \overrightarrow{I_{IJ}^{t,General}} \right]_{(o,p)} \mapsto \left[ \widehat{Co}_{tij}^{(2)} \right]_{(o,p)}. \tag{23}$$

$\widehat{Co}_{ti}^{(2)}$ is the mean predicted concentration by $f^{(2)}$ with respect to the laboratory-prepared reference concentration $Co_{ti}$.

- Average relative error of prediction $REP(\%)$:

$$REP(\%) = \frac{100}{n' \times T} \sum_{i \ in \ S} \sum_{t=1}^{T} \left| \frac{\widehat{Co}_{ti}^{(3)} - Co_{ti}}{Co_{ti}} \right|, \tag{24}$$

$$\widehat{Co}_{ti}^{(3)} = \frac{1}{J''} \sum_{j=1}^{J''} \widehat{Co}_{tij}^{(3)}, \tag{25}$$

Average relative standard deviation $RSD(\%)$ of the predicted concentrations for the validation data set:

$$RSD(\%) = \frac{100}{n'} \sum_{i \ in \ S} \sqrt{\frac{1}{T \times J'' - 1} \sum_{t=1}^{T} \sum_{j=1}^{J''} \left( \frac{\widehat{Co}_{tij}^{(3)} - Co_{ti}}{Co_{ti}} \right)^2}, \tag{26}$$

where $S = \{4\}$ referring to the validation sample set, $\widehat{Co}_{tij}^{(3)}$ is the predicted concentration corresponding to the targeted concentration $Co_{ti}$ by $f^{(3)}$:

$$f^{(3)}: \overrightarrow{I_{IJ}^{t,General}} \mapsto \widehat{Co}_{tij}^{(3)}. \tag{27}$$

- Limit of detection $LOD(\text{ppm})$, deduced by fitting with a straight line, the predicted concentrations $\widehat{Co}_{ijt}$ by $f$ for the calibration sample set:

$$f: \overrightarrow{I_{IJ}^{t,General}} \mapsto \widehat{Co}_{tij}, \tag{28}$$

where $i \in \{1,2,3,5,6,7\}$ referring to the calibration sample set, versus the corresponding prepared concentrations of the calibration sample set, $Co_{ti}$:

$$\widehat{Co}_{ijt} = a + b \times Co_{ti}, \tag{29}$$

$$LOD(\text{ppm}) = \frac{3\sigma_a}{b}, \tag{30}$$

where $\sigma_a$ is the standard deviation of $a$, such variation is due to the dispersion of $\widehat{Co}_{ijt}$. $LOD$ is thus determined by the sensibility of the technique (the slope $b$) and the accuracy and precision of concentration prediction by the model for the different reference samples and different replicates for a given sample (standard deviation of $a$, $\sigma_a$).

In the case of consideration of a specific soil type, the variable $t$ takes the corresponding given value and the concerned sum reduces to a specific term in the above definitions.

**Back-propagation neuronal networks (BPNN).** A single hidden layer BPNN used in this work consists of an input layer, a hidden layer, and an output layer as shown in Fig. 10. The tanh function is used as the activation function of the hidden layer. The Stochastic Gradient Descent (SGD)[42] and Mini-batch Stochastic Gradient Descent (MSGD)[44] iterations are used to construct the BPNN model.

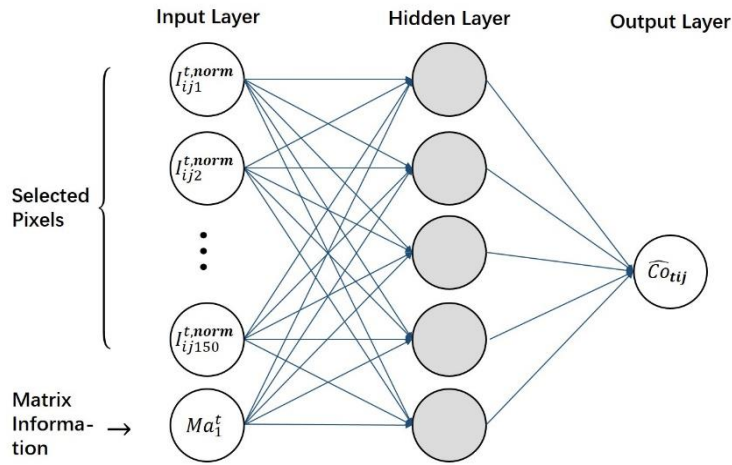

**Figure 10.** Structure of the used neuronal networks.

**Software.** The data processing was carried in the framework of Python version 3.6.4. Scikit-learn and NumPy were used. In addition, Origin Pro 8.0 (Origin Lab Corporation, Northampton, MA, USA) was used to design the figures. All processes were run on a PC (CPU: Intel Core i7-7700 @3.60GHz, RAM: 8.00GB) under Windows 10.

## Acknowledgements

This study was supported by the National Natural Science Foundation of China (Grant Nos. 11574209, 11805126, and 11601327) and the Science and Technology Commission of Shanghai Municipality (Grant No. 15142201000).

## Author contributions statement

C.S. wrote the program and performed the calculations for the multivariate model and participated in the paper writing, Y.T. performed the sample preparation, the experimental measurement and the calculations for the univariate model and participated in the paper writing, J.Y. conceived the experiment and the data treatment strategy, interpreted the results and wrote the paper. N.D.G. participated in the conception of the experimental measurement and the sample preparation, L.G., Y.Z., Z.Y. participated in the data treatment, Y.N., T.Z, H.L. participated in multivariate model construction. All the authors reviewed the manuscript.

## Additional information

**Competing financial interests:** The authors declare no any financial nor non-financial competing interests.